\documentclass[12pt]{article}
\usepackage{graphicx}
\newcommand{\beq}{\begin{equation}}
\newcommand{\eeq}{\end{equation}}
\newcommand{\bea}{\begin{eqnarray}}
\newcommand{\eea}{\end{eqnarray}}

\def\sq{{\vbox {\hrule height 0.6pt\hbox{\vrule width 0.6pt\hskip 3pt
   \vbox{\vskip 6pt}\hskip 3pt \vrule width 0.6pt}\hrule height 0.6pt}}}

\begin{document}
\begin{titlepage}
\begin{flushleft}
       \hfill                      {\tt hep-th/0304051}\\
       \hfill                       FIT HE - 03-02 \\
       \hfill                       KYUSHU-HET 65 \\
\end{flushleft}
\vspace*{3mm}
\begin{center}
{\bf\LARGE Dilaton coupled brane-world and field trapping \\ }
\vspace*{12mm}
{\large Kazuo Ghoroku\footnote[2]{\tt gouroku@dontaku.fit.ac.jp},
Motoi Tachibana\footnote[3]{\tt motoi@postman.riken.go.jp} and
Nobuhiro Uekusa\footnote[4]{\tt uekusa@higgs.phys.kyushu-u.ac.jp}}\\
\vspace*{2mm}

\vspace*{2mm}

\vspace*{4mm}
{\large ${}^{\dagger}$Fukuoka Institute of Technology, Wajiro, 
Higashi-ku}\\
{\large Fukuoka 811-0295, Japan\\}
\vspace*{4mm}
{\large ${}^{\ddagger}$Theoretical Physics Laboratory, RIKEN, 2-1 Hirosawa,}\\
{\large  Wako, Saitama 351-0198, Japan\\}
\vspace*{4mm}
{\large ${}^{\ddagger}$Department of Physics, Kyushu University, Hakozaki,
Higashi-ku}\\
{\large Fukuoka 812-8581, Japan\\}
\vspace*{10mm}
\end{center}

\begin{abstract}
We address solutions of brane-world with cosmological constant $\lambda$
by introducing the dilaton in 5d bulk,
and we examine the localization of graviton, gauge bosons and 
dilaton. 
For those solutions, we find that both graviton and gauge bosons 
can be trapped for either sign, positive or negative, and wide range of 
$\lambda$ due to the non-trivial dilaton. 
While the dilaton can not be trapped on the brane. 
\end{abstract}
\end{titlepage}

\section{Introduction}

The idea of the brane-world given in \cite{RS1,RS2} is strongly
motivated by the recent development in the superstring/M 
theory \cite{HW,MGW}. 
In string theory however, the dilaton is necessarily
included as a scalar partner
of the graviton and plays many important dynamical roles. For example,
the bulk configuration of the dilaton would give an insight for the 
running coupling 
constant of the boundary 4d Yang-Mills theory in an appropriate 10d background 
manifold. It would therefore be interesting to see the role of the dilaton 
in the brane-world and to examine whether the dilaton
could live also in our 4d world after the warped compactification 
and how it behaves. When it survives on the brane 
as a massless scalar partner of the graviton as
in 10d, the equivalence principle, which will be observed in the free-fall
experiments, would be remained as a problem to be resolved as in
\cite{DPV}. While it could appear
as a 4d massive scalar, its mass ($m_{\phi}$) should be very small since it
could not exceed the 4d cosmological constant $\lambda$ in order 
to be trapped \cite{GY}. Then we can 
estimate its magnitude as $m_{\phi}\sim \sqrt{\lambda}$, and the 
coupling to the light particles like photons could be estimated
as $1/M_{\rm pl}$. This implies a very long life time 
$\tau\sim M_{\rm pl}^2/m_{\phi}^3$ of the massive 
dilaton, and this fact gives a serious difficulty known as the moduli 
problem
\cite{RaTh,RT,DV}. In this case, we could however consider this massive 
particle as a candidate of the dark matter \cite{DG,Di}.

It would be meaningful to study the problem of localization of the dilaton
in order to approach to the above issues from the viewpoint of the 
brane-world. Although the localization of
the graviton has been extensively studied \cite{RS2,KR,Mie,KT,bre,BBG},
much attention has not been given to scalar field \cite{NOOO,GN1,GY}, especially
for the dilaton.
However, it would be important to see the situation of the dilaton 
as well as the graviton in the context of the warped compactification
within the superstring theory.

In addition to the dilaton and graviton, we also examine the localization 
of gauge bosons which are the important force of our 4d world. 
Up to now, the localization of the graviton has been
assured in a wide range of solutions except
for AdS${}_4$ brane \cite{KR,Mie}.
As for the gauge bosons, its localization has been examined
in some models \cite{DS,DSG,DFKK,DSL,Oda,KT,GN2,GU}. 
To trap gauge bosons, it was necessary
to introduce positive $\lambda$ or the bulk-dilaton field. 
The latter case might be related to the massive vector model \cite{GN2} by a 
special form of brane-vector coupling \cite{MT}. This point is an 
another reason to examine the brane-world with bulk-dilaton.

\vspace{.2cm}
We study these issues in terms of a simple class of
solutions for brane-world with non-trivial dilaton, which are essentially
equivalent to the one given in \cite{KLO}. However, the use and extension
of these solutions are different.
The dilaton ($\phi$) is identified here with a scalar 
field which couples to the brane
like $e^{\gamma\phi}$ and has the bulk
potential of the exponential form. 
We concentrate on the localization of the fields mentioned above.
Since the solutions given here have the same form of warp factor with the 
one, which are
given before \cite{bre} for the case without the dilaton, then many
problems of the localization would be easily understood. 

\vspace{.3cm}
Several new aspects caused by dilaton are given here.
The localization of both the graviton and 
gauge bosons is realized for any brane of positive, negative and zero
$\lambda$, i.e., dS$_4$, AdS$_4$ and Minkowski brane. 
When dilaton is absent,
the gauge bosons are localized only for $\lambda>0$, dS$_4$ brane \cite{GU}, 
but this
is not a necessary condition when we consider dilaton. 
The essential point needed for the localization of gauge bosons
would be in the deformation of the warp factor from its exact AdS form. 
And the necessary 
deformation can be obtained either by the positive $\lambda$ or by the 
dilaton.
In any case, the conformal symmetry is 
broken, so we could say that
this breaking of the conformal symmetry would be inevitable for
the localization of gauge bosons. 

As for the graviton, it can be localized even if $\lambda<0$ when
dilaton is included. This point is important since the general coordinate
invariance is not necessarily broken for AdS$_4$ brane in the case
when dilaton is considered.

As for the localization of dilaton, we must solve the mixed
equations with the scalar components of the metric. For our solutions,
the master equation for these system is obtained in a simple form, and we
find that any mode of the bulk dilaton can not be trapped on the brane.

\vspace{.2cm}
In Section 2, we set our model with dilaton, and brane solutions
are shown. We comment on the relation between these new solutions and the 
one given before without dilaton. In Section 3, the localization of 
graviton, gauge bosons and dilaton are examined.
In the final section, summary and discussions are given.

\section{Dilaton coupled brane solution}

We begin with the following action with dilaton ($\phi$),
\bea
   S_g=\int d^4\!xdy\sqrt{-g}
   \left\{{1\over 2\kappa^2}(R-2\Lambda) 
    -{1\over 2}(\partial\phi)^2-V(\phi)\right\}  
     -\tau\int d^4\!x\sqrt{-\det g_{\mu\nu}}\ e^{\gamma\phi} \ ,
                                                     \label{acg}
\eea
where the parameters are the five-dimensional gravitational constant 
$\kappa^2$, 
bulk cosmological constant $\Lambda$, 
brane tension $\tau$ and 
dilaton-brane coupling $\gamma$.
The brane is set at $y=0$ by imposing $y\rightarrow -y$ symmetry on 
the above action.
And the form of the potential $V$ is not specified at this stage.
The equations of motion derived from the action (\ref{acg}) are given as
\bea
  &&  G_{MN}=
      \kappa^2\left\{
    \partial_M\phi\partial_N\phi-g_{MN}
    \left({1\over 2}(\partial\phi)^2+{\Lambda\over \kappa^2}+V\right)
    -g_{\mu\nu}\delta_M^{\mu}\delta_N^{\nu}
            \tau e^{\gamma\phi}\delta(y)\right\} \ , \label{eins}
\\
  &&\qquad\qquad
      {1\over\sqrt{-g}}\partial_M\left\{\sqrt{-g}g^{MN}\partial_N\phi\right\}
      ={\partial V\over \partial\phi}
        +{\sqrt{-\det g_{\mu\nu}}\over\sqrt{-g}}
           \tau\gamma e^{\gamma\phi}\delta(y) \ .   \label{dila}
\eea
Here we solve these under the following ansatz for metric, 
\bea
   ds^2=g_{MN}dx^M dx^N=A^2(y)(-dt^2+a^2(t)\gamma_{ij}dx^i dx^j)
         +dy^2 \ ,                                 \label{fmet}
\eea
where $\gamma_{ij}=(1+k\delta_{mn}x^m x^n/4)^{-2}\delta_{ij}$.
As long as we do not mention, $k=0$.
While $a(t)$ is solved for each $k$, for example
$a_0=e^{H_0 t}$, $H_0=\sqrt{\lambda}$ for $k=0$.
If we take as $\phi=\phi(y)$, 
Einstein equations (\ref{eins}) and dilaton equation (\ref{dila}) are 
written as
\bea
   &&{A''\over A}+\left({A'\over A}\right)^2
     -{\lambda \over A^2}
  =-{\kappa^2\over 3}
       \left({1\over 2}(\phi')^2 +{\Lambda\over \kappa^2}+V(\phi)\right)
     -{\kappa^2\tau\over 3}e^{\gamma\phi}\delta(y) \ ,             
\label{tteq}
\\
    &&\qquad\qquad
      \left({A'\over A}\right)^2-{\lambda \over A^2} 
    ={\kappa^2\over 6}
     \left({1\over 2}(\phi')^2 -{\Lambda\over \kappa^2}-V(\phi)\right) \ ,
                                                  \label{yyeq}
\\
    &&\qquad\qquad
          \phi''+4\phi'{A'\over A}
        ={\partial V\over \partial \phi}
         +\tau\gamma e^{\gamma\phi}\delta(y) \ ,  \label{deq}
\eea
where $'=d/dy$.
Let us assume the following form for $\phi$,
\bea
    \phi'=\alpha A^{n}\rm{sgn}(y)\ ,   \label{ans}
\eea
Eqs.(\ref{yyeq}) and (\ref{deq}) are solved as
\bea
    &&\qquad
      \left({A'\over A}\right)^2-{\lambda \over A^2} 
    =-{\kappa^2\over 6}
     \left({2\alpha^2\over n}A^{2n} + V_0\right) -{\Lambda\over 6}\ ,
                                                  \label{yyeq1}
\\
    &&\qquad
     V= {n+4\over 2n}({\alpha A^n})^2+V_0, \qquad 
           2\alpha =\gamma\tau e^{\gamma\phi_0}\ ,
          \label{deq1}
\eea
where $V_0$ is a constant, and we take the boundary conditions $A(0)=1$ and 
$\phi(0)=\phi_0$.
From eq.(\ref{yyeq1}), we can see that the parameters $V_0$ and $\alpha$
can be absorbed into $\Lambda$ and $\lambda$ when $n=-1$. We can solve for 
other
values of $n$, but we consider this simple solution in the following 
analysis
since it is included in the
previously obtained one for the case without dilaton by replacing the 
modified parameters. We should however notice here
that this solution of $n=-1$ corresponds to the one which has 
already been found by P.~Kanti 
{et al.}\cite{KLO}. Another ansatz which could lead to the solutions
similar to our previous ones are also possible, but they are a little bit
complicated. So we concentrate here on the solution of $n=-1$ with the above
ansatz (\ref{ans}).

\vspace{.5cm}
Then the potential is given as
\bea
   V= -{3\over 2}\biggl({\alpha\over A}\biggr)^2+V_0,    \label{poten}
\eea
and the equations (\ref{tteq}) and (\ref{yyeq}) can be written as
\bea 
    &&
   {A''\over A}+\left({A'\over A}\right)^2
     -{\tilde{\lambda}\over A^2}+{\tilde{\Lambda}\over 3}
    =-{\kappa^2\tau\over 3}e^{\gamma\phi_0}\delta(y) \ .        
\label{tteq3}
\\   &&\qquad
     \left({A'\over A}\right)^2+{\tilde{\Lambda}\over 6}
    ={\tilde{\lambda}\over A^2} \ ,        \label{yyeq2}
\\
    &&
       \tilde{\Lambda}=\Lambda+\kappa^2V_0 \ , \quad
      \tilde{\lambda}=\lambda+{\kappa^2\alpha^2\over 3} \ . \label{lam}
\eea
As stated above,
the equations (\ref{tteq3}) and (\ref{yyeq2}) are the same form with the 
one
obtained before \cite{bre} for the case where dilaton is suppressed. 
The dilaton modifies superficially the brane and bulk cosmological 
constants
($\lambda, \Lambda$) to $(\tilde{\lambda}, \tilde{\Lambda})$ as shown 
above.
So the
solutions are obtained by replacing the parameters ($\lambda, \Lambda$) 
in the previously given solutions by the new one, 
$(\tilde{\lambda}, \tilde{\Lambda})$. 
Then the various analyses can be
performed in a parallel way.
For example,
the solutions $A(y)$ and $\phi(y)$ and the potential $V(\phi)$ are given as
follows.

\vspace{.4cm}
\noindent For $\tilde{\lambda}>0,\ \tilde{\Lambda}=0$,
\bea
   &&\qquad\qquad
    A(y)=1-\sqrt{\tilde{\lambda}}|y|\equiv 1-{|y|\over y_H} \ 
,\label{singsol}
\\
   &&\phi(y)=-{\alpha\over\sqrt{\tilde{\lambda}}}
           \ln(1-\sqrt{\tilde{\lambda}}|y|)+\phi_0 \ ,\quad
   V(\phi)=-{3\alpha^2\over 2}e^{{2\sqrt{\tilde{\lambda}}\over\alpha}
                   (\phi-\phi_0)}
       +V_0 \ .                     \label{scalarp1}
\eea
For $\tilde{\lambda}>0,\ \tilde{\Lambda}<0$,
\beq
       A(y)={\sqrt{\tilde{\lambda}}\over \mu}\sinh\left[\mu(y_H-|y|)\right] 
\ ,
\eeq
\beq
     \phi(y)={\alpha\over \sqrt{\tilde{\lambda}}}
           \ln\left({\coth\left[\mu(y_H-|y|)/2\right]\over
                     \coth(\mu y_H/2)}\right)+\phi_0 \ ,
\eeq
\beq
     V(\phi)=-{3\alpha^2\over 2}{\mu^2\over \tilde{\lambda}}
              \sinh^2\left\{\sqrt{\tilde{\lambda}}(\phi-\phi_0)/ \alpha
             +\chi\right\}
              +V_0 \ ,            \label{scalarp2}
\eeq
\beq
     \mu=\sqrt{-\tilde{\Lambda}/6} \ ,\quad
      \sinh(\mu y_H)=\mu/\sqrt{\tilde{\lambda}} \ ,\quad
     e^{\chi}= \coth(\mu y_H/2) \ .
\eeq
For $\tilde{\lambda}>0,\ \tilde{\Lambda}>0$,
\beq
    A(y)={\sqrt{\tilde{\lambda}}\over \mu_d}\sin\left[\mu_d(y_H-|y|)\right] 
\ ,
     \phi(y)={\alpha\over \sqrt{\tilde{\lambda}}}
           \ln\left({\cot\left[\mu_d(y_H-|y|)/2\right]\over
                    \cot(\mu_d y_H/2)}\right)+\phi_0\ ,\quad
\eeq
\beq
     V(\phi)=-{3\alpha^2\over 2}{\mu_d^2\over \tilde{\lambda}}
            \cosh^2\left\{\sqrt{\tilde{\lambda}}(\phi-\phi_0)/\alpha
            +\chi_d\right\}
              +V_0 \ ,                     \label{scalarp3}
\eeq
\beq
     \mu_d=\sqrt{\tilde{\Lambda}/6} \ ,\quad
      \sin(\mu_d y_H)=\mu_d/\sqrt{\tilde{\lambda}} \ ,\quad 
       e^{\chi_d}=\cot(\mu_d y_H/2) \ .
\eeq
For any above solution,
there exists horizon whose position is represented by $y=y_H$
and which has curvature singularity 
due to the non-trivial dilaton.
So we interpret the extra dimension is in a finite region,
$0\le y\le y_H$. 
Hereafter we set $\phi_0=0$ for the sake of brevity, or we interpret
$\tau$ as $\tau e^{\gamma\phi_0}$ as long as we do not mention it especially.
And we obtain from eq.(\ref{tteq3}) for the above solutions
\bea
      \tilde{\lambda}={\tilde{\Lambda}\over 6}+
              {\kappa^4\tau^2\over 36} \ .        \label{lambda}
\eea
This is the same form of the relation given before by $\Lambda$ and 
$\lambda$ in the case
of the theory without dilaton.

\section{Localization}

Here we consider the localization of the bulk fields in terms of the
linearized field equations for the graviton, dilaton and gauge bosons
around the background configuration obtained above.

\subsection{Graviton} 

Under the metric (\ref{fmet}), it is convenient to take the gravitational 
fluctuation $h_{ij}$ as follows,
\beq
 ds^2= A(y)^2(-dt^2+a^2(t)[\gamma_{ij}(x^i)+h_{ij}(x^{\mu},y)]dx^{i}dx^{j})
           +dy^2  \,. \label{metricape}
\eeq
We are interested in the localization of the traceless transverse
component, which represents the graviton on the brane.
It is projected out by the conditions, $h_i^i=0$ and
$\nabla_i h^{ij}=0$, where $\nabla_i$ denotes the covariant derivative
with respect to the three-metric $\gamma_{ij}$ which is used to raise
and lower the three-indices $i,j$. The transverse and traceless part
is denoted by $h$ hereafter for simplicity. The linearized field equation
for $h$ is obtained in the same form with the one given previously for
the case without dilaton,
\beq
 \nabla^2_5h=0,  \label{scalar}
\eeq
in terms of
the five dimensional covariant derivative $\nabla^2_5=\nabla_M\nabla^M$.
The above equation is given for $k=0$, and
this is equivalent to the field equation of a five dimensional 
free massless scalar. For $k=\pm 1$, the term proportion to this appears,
but it is not essential here and is abbreviated.

Then we arrive at the conclusion that the graviton can be localized on
the brane if $\tilde{\lambda}\geq 0$. The new point is that the graviton
can be trapped even if $\lambda <0$ differently from the case without
dilaton. When the dialton is not considered, massive spin-2 field is 
trapped
on the brane of negative $\lambda$. Then the general coordinate invariance
is broken for AdS$_4$ brane from the viewpoint of the brane world scenario.
But this is not always true as seen above.

For scalar fields, we can obtain similar results given for the case
without dilaton by replacing the parameters $(\Lambda,\lambda)$ by
$(\tilde{\Lambda},\tilde{\lambda})$. The discussions are abbreviated here.

\subsection{Dilaton} 

Here we discuss the dilaton according to its linearized equation
around the background solution given here. The equation is obtained
by denoting as $\phi=\bar{\phi}+\delta\phi$, where $\bar{\phi}$
represents the classical solution for $\phi$.
The bulk ``mass'' term of the dilaton is obtained from 
the potential, $V(\bar{\phi}+\delta\phi)$ given above. 
Expanding it around the classical solutions,
we get
\beq
 V=\bar{V}+{1\over 2}m_{\phi}^2(\delta{\phi})^2,
\eeq
\beq
   m_{\phi}^2=-{6\tilde{\lambda}\over A^2}+{\tilde{\Lambda}\over 2},
\eeq
where $\bar{V}=V(\bar{\phi})$. 
The explicit forms of $m_{\phi}^2$ for each solutions
are written as
\bea
   &&\qquad 
     m_{\phi}^2=-6\tilde{\lambda}
      e^{{2\sqrt{\tilde{\lambda}}\over\alpha}\bar{\phi}} \ ,
     \qquad\qquad\qquad\  \textrm{for } \tilde{\Lambda}=0,\ 
\tilde{\lambda}>0 ,
\\
  && m_{\phi}^2=-3\mu^2\cosh\left[\right.2
        ({\sqrt{\tilde{\lambda}}\over \alpha}\bar{\phi}+\chi)\left.\right] 
\ ,
     \qquad \textrm{for } \tilde{\Lambda}<0, \ \tilde{\lambda}>0 ,
\\
  && m_{\phi}^2=-3\mu_d^2\cosh\left[\right.2
       ({\sqrt{\tilde{\lambda}}\over \alpha}\bar{\phi}+\chi_d)\left.\right] 
\ ,
     \qquad \textrm{for } \tilde{\Lambda}>0,\ \tilde{\lambda}>0 .
\eea 
They are all not constant
but the functions of $y$. In this sense, we can not see the mass of
the dilaton in a normal form. On the other hand, we can see that they are all
negative in all region of $y$. So the brane obtained here seems to be
unstable for the dilaton fluctuation. However, we 
should notice the other mass 
term coming from the brane, and the total mass term is given as
\beq
   m_{\phi}^2=-{6\tilde{\lambda}\over A^2}+{\tilde{\Lambda}\over 2}
             +\tau\gamma^2\delta(y).  \label{dilaton-mass}
\eeq
We can see a possibility to cure the instability by the third term. 
In order to 
make clear this point we must however solve the coupled equations with the 
scalar component of the metric and dilaton fluctuations.

\vspace{.5cm}
In order to reduce the number of the scalar components of the metric,
we move to the conformal coordinate by
changing the fifth coordinate $y$ to $z$ in terms of the relation,
$dy/dz=A(y)$. Then the metric is rewritten as
\bea
   ds^2=g_{MN}dx^M dx^N=A^2(z)(\gamma_{\mu\nu}dx^{\mu} dx^{\nu}
         +dz^2) \ ,                                 \label{fmet2}
\eea
where
\bea
   \gamma_{\mu\nu}dx^{\mu} dx^{\nu}=
     (-dt^2+a^2(t)\gamma_{ij}dx^i dx^j) \ .  \label{fmet3}
\eea
In this coordinate, the scalar fluctuations of the metric are reduced to
the following form in an appropriate gauge \cite{KKS,Gio},
\begin{equation}
ds^2 = A^2(z)\left[(1+2\sigma)dz^2
+(1+2\psi)\gamma_{\mu\nu}dx^{\mu}dx^{\nu}\right]. \label{scalar-fluctuation}
\end{equation}

\vspace{.3cm}
After a calculation, we obtain the linearized equation for $\psi$,
\beq
-\partial_z^2{\psi}-3\mathcal{H}\partial_z{\psi}
-\left(4\partial_z{\mathcal{H}}+6\lambda\right)\psi=\sq_4\psi
   \, , \label{psi-equation}
\eeq
where $\mathcal{H}=\partial_z{A}/A$ and $\sq_4$ denotes the four
dimensional laplacian as given below. In deriving this
equation, the constraints
\beq
 \sigma=-2\psi, \qquad \delta\phi\partial_z\bar{\phi}=
     -3(\partial_z{\psi}+2\mathcal{H}{\psi})
\eeq
and classical equations are used. In this process, the dilaton-brane
coupling $\gamma$ is disappeared in (\ref{psi-equation})
due to a characteristic property of our classical
solution for the dilaton, the case of $n=-1$ in (\ref{ans}).
Then, $\psi$ is decomposed as follows
in terms of the four-dimensional continuous mass eigenstates:
\beq
 \psi=\int\! dm \ \varphi_m(t,x^i)\,\Phi(m,y) \, , \label{eigenex}
\eeq
where the 4d mass $m$ is defined by
\beq
 -\sq_4\varphi=\ddot{\varphi}_m+3{\dot{a}_0\over a_0}\dot{\varphi}_m
           +{-\partial_i^2\over a_0^2}\varphi_m=-m^2\varphi_m , 
\label{masseig}
\eeq
and $\dot{}=d{}/dt$. In order to see the localization, we do not
need the solution of this equation, and
we need only the explicit form
of $\Phi(m,y)$. Its equation is rewritten 
into the form of one-dimensional 
Schr\"{o}dinger-like equation with the rescaled $\Phi(m,y)$ as 
$\Phi(m,y)=A^{-3/2}u(z)$ and
modified eigenvalue $\tilde{m}^2$,
\beq
 [-\partial_z^2+V(z)]u(z)=\tilde{m}^2 u(z) , \ \label{warp3}
\eeq
where
\beq
  \tilde{m}^2=m^2+6{\lambda} \, ,\qquad 
\eeq
\beq
 V(z)={9\over 4}(A')^2-{5\over 2}AA''.  \label{potential-psi}
\eeq 
where we must notice ${}'=d{}/dy$.
Since $A$ is singular at $y=0$ ( at $z=z_0$), then $u$ must satisfy the 
following boundary condition
at $z=z_0$,
\beq
 u'(z_0)={5\kappa^2\tau\over 12}u(z_0).  
                   \label{boundzero1}
\eeq
And the eigenvalue $\tilde{m}^2$ of the bound state is given as the 
solution
of this equation (\ref{boundzero1}).

\vspace{.2cm}
Without solving this equation, we can say that any state of $u(z)$
can not be trapped on the brane due to the positive $\delta$-function potential
included in the second term of the above potential (\ref{potential-psi}). 
We must notice that
the potential becomes negative in some region for our background solutions,
and this implies a tachyonic state could be trapped on the brane.
But it is not a problem here since such a dangerous mode can not be trapped
as stated above due to the property of the potential. So 
we can say that the brane considered
here is stable for the dilaton fluctuation.

\subsection{Gauge field}

Here
we consider a case where the dilaton couples to gauge fields
with the following action
\bea 
   &&  S=S_g+S_{\textrm{\scriptsize gauge}} \ ,   \label{action}
\\
   &&  S_{\textrm{\scriptsize gauge}}=
           \int d^4x \int dy \sqrt{-g}
             \left(-{1\over 4}e^{4\zeta\phi}g^{MN}g^{PQ}F_{MP}F_{NQ}\right) 
\ ,
                                                    \label{actiond}
\eea
where $\zeta$ denotes the dilaton-gauge field coupling.
When the dilaton is decoupled from the gauge fields, then $\zeta=0$ and
the gauge fields can be trapped for $\tilde{\lambda}>0$ in this case. 
This is easily 
understood from the previous analysis \cite{GU} given for the case 
without dilaton. In fact, the warp 
factor and the equation of the gauge-boson fluctuation have the same form 
with the one given in \cite{GU} if ($\tilde{\lambda}$, $\tilde{\Lambda}$) 
were replaced by ($\lambda$, $\Lambda$).

\vspace{.2cm}
For $\zeta\neq 0$,
we expand the fields in terms of the four-dimensional mass eigenstates,
$A_M=\int dm a_M(t,x^i,m)\phi_M(y,m)$, and 
the equation of motion for the spatial transverse
component $A_i^T$
is derived as
\bea
  &&  (\partial_y^2
   +\{2{A'\over A}+4\zeta\phi'\}\partial_y
   +{m^2\over A^2})\phi_i^T           
    =0 \ ,                                 \label{atrans}
\\
  &&\qquad
     (-\partial_t^2-{\dot{a}_0\over a_0}\partial_t+{\partial_j^2\over 
a_0^2}
     )a_i^T=m^2 a_i^T \ .              
\eea
By introducing $u(z)$ and $z$ defined as 
$\phi_i^T=A^{-1/2}u(z)$ and 
$\partial z/\partial y=\pm A^{-1}$, Eq.(\ref{atrans}) can be written as 
\beq
 [-\partial_z^2-4\xi\partial_z+V(z)]u(z)=m^2 u(z) , \ \label{warp31}
\eeq
where $\xi=\zeta\alpha$ and the potential $V(z)$ is given as
\beq
 V(z)={1\over 4}(A')^2+{1\over 2}AA''+2\xi A' .  \label{pott}
\eeq 
%
It is difficult to get an analytic solution
of the equation (\ref{warp31}) for non-zero $m$
except for some special cases, which
are given in two examples below. Before showing them, we show that
the zero mode is trapped on the brane for all solutions given above.
In other words, the gauge bosons can be trapped for those solutions
when the dilaton couples to the gauge bosons. It is proved as follows.

\vspace{.3cm}
For $\zeta=0$, the eigenvalue equation is written as
\beq
 [-\partial_z^2+V_0(z)]u(z)=m^2 u(z) , \ \label{warp4}
\eeq
\beq
 V_0(z)={1\over 4}(A')^2+{1\over 2}AA'' ,
\eeq 
For the zero-mode $u_0(z)$ ($m=0$), Eq.(\ref{warp4}) is written as, 
\beq
 [-\partial_z^2+V_0(z)]u_0(z)=0 , \ \label{warp40}
\eeq
and its normalizable solution is obtained as
\beq
  u_0(z)=c_0 X^{1/4}                 \label{solution}
\eeq
where $c_0$ is a constant. And $X={\rm sinh}^{-2}(\sqrt{\tilde{\lambda}}z)$
($X={\rm cosh}^{-2}(\sqrt{\tilde{\lambda}}z)$) for ${\tilde{\lambda}}>0$
and ${\tilde{\Lambda}}<0$ (${\tilde{\Lambda}}>0$). 

\vspace{.3cm}
It is easily seen that the above solution $u_0(z)$ also satisfies the zero-eigenvalue equation of $\zeta\neq 0$,
\beq
 [-\partial_z^2-4\xi\partial_z+V(z)]u_0(z)=0 , \ \label{warp310}
\eeq
where $\xi=\zeta\alpha$ and the potential $V(z)$ is given as
\beq
 V(z)=V_0(z)+2\xi A' .  \label{pott01}
\eeq 
The proof of (\ref{warp310}) is shown as follows.
By substituting the solution (\ref{solution}) into (\ref{warp310}),
we find
\beq
 [-\partial_z^2-4\xi\partial_z+V(z)]u_0(z)=2\xi(-2\partial_z+A')u_0(z)
 , \ \label{warp311}
\eeq
and the right hand side vanishes when the explicit forms of $u_0(z)$
and $A'(z)$ for each solution given above are used. Then the
zero-mode solutions are not affected by the dilaton coupling.

\vspace{.3cm}
On the other hand, 
the normalizability condition is modified by the dilaton coupling.
When dilaton couples to the gauge bosons, the condition of normalizability 
of zero mode is given as
\beq
  \int_{z_0}^{\infty}dz e^{4\xi z}u_0(z)^2 < \infty , \label{normali}
\eeq
From this and the explicit form of $u_0(z)$, 
we find for both solutions the following common condition,
\beq
  4\xi < \sqrt{\tilde{\lambda}} . \label{const1}
\eeq

\vspace{.5cm}
As a result, we can say 
that the gauge bosons can be trapped for the case of $\zeta\neq 0$ whenever the constraint (\ref{const1})
is satisfied.
In the followings, we can see this relation through the explicit solutions
for the soluble cases for general $m$.

\vspace{.5cm}
As a first example, we consider the case of $\tilde{\lambda}>0$ and
$\tilde{\Lambda}=0$. This is given as a limit of $\tilde{\Lambda}\to 0$
from either solution of $\tilde{\Lambda}>0$ or $\tilde{\Lambda}<0$. 
In this case, we can solve for $u(z)$ of non-zero $m$, but
the setting of
$\tilde{\Lambda}=0$ would not be realistic when we respect the Newton's law
to be observed in our world. Although the graviton is trapped on the brane,
its KK modes would be overwhelming \cite{GNY}. 
So the analysis of this case is 
performed from the theoretical interest.

As above the same form of equation
(\ref{warp31}) and (\ref{pott}) are obtained. And the explicit form
of the potential is given as
\beq
 V(z)={\sqrt{\tilde{\lambda}}\over 4}(\sqrt{\tilde{\lambda}}-8\xi)
            -\sqrt{\tilde{\lambda}}\delta(z-z_0) .  \label{pott0L}
\eeq 
Then $u(z)$ is solved as,
\bea
 u(z)=c_1e^{k_+z}+c_2e^{k_-z},  \qquad 
         k_{\pm}=-2\xi \pm\sqrt{(2\xi-\sqrt{\tilde{\lambda}}/2)^2-m^2}
\label{soldsL}
\eea
where $c_1$ and $c_2$ are constants of integration. The normalizable 
solution is obtained by choosing $k_-$ and the boundary condition,
(\ref{boundzero}), can be written as
\beq
 -2\xi+\sqrt{\tilde{\lambda}}/2=\sqrt{(2\xi-\sqrt{\tilde{\lambda}}/2)^2-m^2}\ .
                           \label{abovc} 
\eeq
Then we obtain $m=0$ and $\xi<\sqrt{\tilde{\lambda}}/4$, which is nothing 
but the one given above, (\ref{const1}). This indicates that the result given
above for the zero-mode is satisfied also 
in the limit of $\tilde{\Lambda}\to 0$.

\vspace{.3cm}
As a second soluble example, we 
consider another limit, $\tilde{\lambda}=0$. However,
$\lambda$ is zero or negative in this limit as seen from the relation 
$\tilde{\lambda}=\lambda+{\kappa^2\alpha^2\over 3}$ (see (\ref{lam})), so
the brane is not a realistic also in this case 
when we respect the present observation
of small but positive $\lambda$. 
However it is meaningful from a theoretical viewpoint to study this case. 

In this case,
the warp factor $A(y)$ takes the simple Randall-Sundrum form, and
the potential is given as
\beq
 V(z)={3\over 4z^2}-{2\xi\over z}-\mu\delta(|z|-z_0) .  \label{pott0}
\eeq 
Then $u(z)$ is solved as,
\bea
 u(z)=e^{-2z(\xi+d/2)}z^{3/2}\{c_1~{}_1F_1(b_1,b_2;2zd)
+c_2 U(b_1,b_2;2zd)\}, 
        \label{solds}
\eea
where $c_1$ and $c_2$ are constants of integration and
\beq
    d=\sqrt{4\xi^2-m^2},\qquad b_1={3\over 2}-{\xi\over d},
          \qquad b_2=3     \label{para1}
\eeq
Here ${}_1F_1(b_1,b_2;X)$ denotes the Kummer's hypergeometric function,
and the confluent hypergeometric function denoted by
$U(b_1,b_2;X)$ is another independent solution of the same differential
equation.
It follows from this solution that $u(z)$ oscillates 
with $z$ when $m^2>4\xi^2$, 
where the continuum KK modes appear. 

\vspace{.5cm}
Here we concentrate on the bound state which is restricted to the region
of $m^2<4\xi^2$. From the normalization condition, 
which is obtained by replacing $u_0(z)$ in (\ref{normali}) by $u(z)$,
\beq
  \int~e^{4\xi z}u^2(z)dz <\infty,
\eeq
the solution for the localized state is obtained
by setting $c_1=0$ since 
$$e^{2\xi z}~e^{-2z(\xi+d/2)}z^{3/2}{}_1F_1(b_1,b_2;2zd)\to e^{2|\xi| 
z}/\sqrt{z}$$ for 
$z\to \infty$. Then $u(z)$ is written as,
\beq
 u(z)=c_2e^{-2z(\xi+d/2)}z^{3/2}U(b_1,b_2;2zd).
 \label{solb}
\eeq

At the next step, we consider the boundary condition at $z=z_0$ (or $y=0$), which
is given as follows by taking into account of the 
$\delta$-function potential in (\ref{pott0}),
\beq
{du(z_0)\over dz}=-{1\over 2}({\kappa^2\tau\over 6})u(z_0).  
\label{boundzero}
\eeq
This condition can be written as
\beq
  {1\over 2}+{\mu-\xi\over d}={U(b_1,b_2+1;2d/\mu)\over U(b_1,b_2;2d/\mu)},
   \label{bound03}
\eeq
where we used $z_0=1/\mu$. This equation provides the value of $m$ when
$\xi$ and $\mu$ are given. Our interest is in the trapped gauge bosons, so
we see this equation for $m=0$. For $\xi>0$, the above equation is
written as
\beq
  {2\over x}={U(1,4;x)\over U(1,3;x)},
   \label{bound00}
\eeq
where $x=4|\xi|/\mu$. While the right hand side is obtained as 
\footnote[4]{
The relations (\ref{bound02}) and (\ref{bound04}) are easily obtained 
from the identities, $U(a,a+1;x)=x^{-a}$ and 
$U'(a,c;x)=U(a,c;x)-U(a,c+1;x)$}
\beq
  {U(1,4;x)\over U(1,3;x)}={2\over x}+{x\over 1+x},
   \label{bound02}
\eeq
and we can see (\ref{bound00}) is not satisfied for $x>0$. 
From this we can say that there is no solution of $m=0$ for $\xi>0$.

\begin{figure}[htbp]
\begin{center}
\voffset=15cm
  \includegraphics[width=9cm,height=7cm]{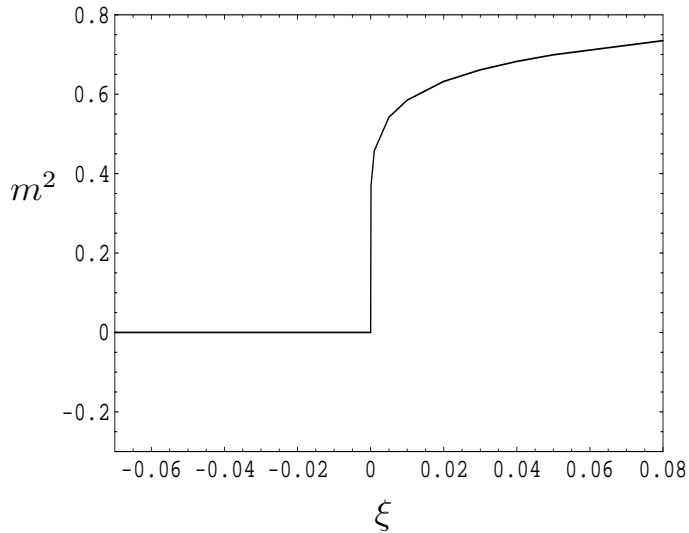} 
\caption{The value $m^2$ (vertical-axis) of the solution of equation 
(\ref{boundzero}) versus $\xi$ (horizontal-axis) for $\mu=1$.
\label{mxigraph2}}
\end{center}
\end{figure}

For $\xi<0$, equation (\ref{bound03}) is written as
\beq
  1+{2\over x}={U(2,4;x)\over U(2,3;x)}.
   \label{bound04}
\eeq
This equation is the identity, so we find always the solution $m=0$
for $\xi< 0$. As a result, we can say that the gauge bosons can be
trapped in the region, $\xi< 0$,
for $\tilde{\lambda}=0$. This result is again consistent with the
relation $4\xi<\sqrt{\tilde{\lambda}}$, (\ref{const1}).
In order to see clearly the above result, the numerical estimation
of the value of $m$ as the solutions of (\ref{boundzero}) is shown
in the Fig.\ref{mxigraph2} as a function of $\xi$ for fixed $\mu$, $\mu=1$.

\vspace{.2cm}
For the case of $\tilde{\lambda}<0$,
the graviton can not be trapped as stated above, then this brane 
is also unrealistic. While 
it would be worthwhile from the theoretical viewpoint
to study this case, but we like to discuss in the future.

\vspace{.4cm}
Finally at this sub-section, we comment on another interesting point
stated in the introduction for the case of nonzero $\zeta$.
The dilaton-coupled action (\ref{action}) could be
related to the action described in terms of massive vector field.
According to \cite{MT}, where the case of zero effective cosmological constant 
$\lambda$ is considered,
we define the massive vector field as
$\tilde{A}_M \equiv \epsilon A_M$ and $\epsilon\equiv e^{2\zeta\phi}$.  
Then the mass-like term for the component $\tilde{A}_{\mu}$
is given by
\beq
    -\left({A'\over A}{\epsilon'\over \epsilon}+
           {1\over 2} \left( {\epsilon'\over \epsilon}\right)'
                                  +\frac{1}{2}\left({\epsilon'\over 
\epsilon}\right)^2\right)
            \sqrt{-g}g^{\mu\nu}\tilde{A}_{\mu}\tilde{A}_{\nu} \ ,
\label{gaugemass}
\eeq
where the delta function term is also included.
For the brane solutions given above,
the ``mass'' is dependent on $y$ in general, 
then
it can not be identified as the standard mass. 
This 
would imply that we need special brane solution to rewrite
the theory to the massive gauge theory with a special coupling to the 
brane.
On this point, we will discuss in a separate paper.


\section{Summary and Discussions}

We have examined the solutions of brane-world including the 
dilaton
and investigated the localization of the bulk fields. 
Due to a specific ansatz, 
the solutions are obtained in a way where
only the brane and bulk cosmological constant  
$(\lambda,\Lambda)$ are modified to $(\tilde{\lambda},\tilde{\Lambda})$
without changing the equations of motion compared to the case without
dilaton field. The new point is the appearance of a dynamical bulk 
scalar, the dilaton.

The solutions are classified by these new parameters 
$(\tilde{\lambda},\tilde{\Lambda})$. For $\tilde{\lambda}>0$, the graviton
and the gauge bosons are trapped as shown previously. 
The new feature is seen from $\tilde{\lambda}=\lambda+\kappa^2\alpha^2/3$.
When dilaton is decoupled ($\alpha=0$), the graviton is trapped only 
for $\lambda\geq 0$, but 
it is trapped even if $\lambda<0$ when the dilaton couples and
$\tilde{\lambda}>0$ were satisfied. This is always possible since 
$\kappa^2\alpha^2/3>0$. This point is important in the sense that we can 
see
the graviton also in AdS$_4$ brane. The breaking of the general coordinate
invariance is expected in AdS$_4$ brane, but this is evaded by 
including the dilaton in such a way to satisfy $\tilde{\lambda}>0$. 
We should however notice that this breaking would be seen
in AdS$_4$ brane when $\lambda$ is very small and 
we arrive at $\tilde{\lambda}<0$. 

As for the gauge bosons, the new analyses are given here by including
the coupling ($\zeta$) of the dilaton to gauge bosons. We find that
the gauge bosons are trapped for small enough
coupling $4\xi(=4\zeta\alpha)<\sqrt{\tilde{\lambda}}$ (see (\ref{const1})).
This condition is further assured by studying two concrete examples, where
solutions for any 4d mass-state could be obtained.

When we consider the smallness of $\lambda$ and 
$\tilde{\lambda}=\lambda+(\kappa\alpha)^2/3$, the above condition is
satisfied if $\zeta$ is the same order of the 5d gravitational coupling
$\kappa$ or smaller than it. As a limit of this constraint, we can consider
the case of $\zeta=0$, the decoupling limit. Our results given here
are consistent with the one given in the previous analysis for
$\zeta=0$ where the trapping of the gauge bosons is observed for
small $\lambda$.
One more point to be noticed is the relation between the dilaton coupled
gauge theory and the massive vector theory. 
We will address on this point in near future.

Finally we comment on the localization of the dilaton which 
should be considered as an important dynamical field as well as graviton
from the string theory viewpoint. 
At first glance, there seems a possible tachyonic bound state of
the dilaton fluctuations since its bulk potential is negative.
The trapping problem of this field however should be 
solved as a mixed system with scalar components of
the metric fluctuations. We could then find a simple master equation
for this system and that there is no trapping of any mode of
dilaton fluctuations and also for the scalar components of the metric.
As a result, we obtain a stable brane model with a dilaton.
The massless dilaton is 
usually expected as in the 10d superstring theory, but our analysis 
implies that the dilaton is absent
in our 4d universe after a compactification. 
From the viewpoint of superstring theory, this seems to be
a difficulty at a glance, but it
could give a possible resolution for the moduli problem and the equivalence
principle when the dilaton is rejected to exist in our 4d world.

The bulk dilaton can be
related to the problem of the AdS/CFT correspondence. When the dilaton
is trivial or absent, the bulk is AdS$_5$. And it has been
observed that the propagator
of the graviton on the brane seems to receive quantum corrections from 
CFT living on the same brane \cite{DL,Gidd}. This correction is
seen through the Newton potential between two massive objects on the brane, 
and the same correction is obtained
from the 5d propagator of the graviton in the AdS bulk.
This is known as AdS/CFT correspondence
which is consistent with the conformal invariance. 
For the case of the non-trivial dilaton, we 
expect a non-conformal field theory on the boundary.
So it would be an interesting problem to study this problem
in our model with non-trivial dilaton configurations.
A report on these points would be given in a future paper.

\vspace{.3cm}
\section*{Acknowledgments}
This work has been supported in part by the Grants-in-Aid for
Scientific Research (13135223)
of the Ministry of Education, Science, Sports, and Culture of Japan.
M.T. was supported by Special Postdoctral Research
Program in RIKEN.


\end{document}